\documentclass[english,pre,twocolumn,showpacs]{revtex4}
\usepackage[T1]{fontenc}
\usepackage[latin1]{inputenc}
\usepackage{amssymb}

\makeatletter

\usepackage{babel}
\makeatother
\begin{document}
\mbox{\hspace{15.8cm}ECT*-04-07}

\title{Forward-backward equations for nonlinear propagation in axially-invariant
optical systems}

\author{Albert Ferrando and Mario Zacarés.}

\affiliation{Departament d'Òptica, Universitat de València. Dr. Moliner, 50. E-46100
Burjassot (València), Spain.}

\author{Pedro Fernández de Córdoba. }

\affiliation{Departamento de Matemática Aplicada, Universidad Politécnica de Valencia.
Camino de Vera, s/n. E-46022 Valencia, Spain.}

\author{Daniele Binosi.}

\affiliation{ECT{*}, Villa Tambosi, Strada delle Tabarelle 286, I-38050 Villazzano
(Trento), Italy.}

\author{Álvaro Montero. }

\affiliation{Mediterranean University of Science and Technology. Camino de Vera,
s/n. E-46022 Valencia, Spain.}

\date{\today}

\begin{abstract}
We present a novel general framework to deal with forward and backward
components of the electromagnetic field in axially-invariant nonlinear
optical systems, which include those having any type of linear or
nonlinear transverse inhomogeneities. With a minimum amount of approximations,
we obtain a system of two first-order equations for forward and backward
components explicitly showing the nonlinear couplings among them.
The modal approach used allows for an effective reduction of the dimensionality
of the original problem from $3+1$ (three spatial dimensions plus
one time dimension) to $1+1$ (one spatial dimension plus one frequency
dimension). The new equations can be written in a spinor Dirac-like
form, out of which conserved quantities can be calculated in an elegant
manner. Finally, these new equations inherently incorporate spatio-temporal
couplings, so that they can be easily particularized to deal with
purely temporal or purely spatial effects. Nonlinear forward pulse
propagation and non-paraxial evolution of spatial structures are analyzed
as examples. 
\end{abstract}

\pacs{42.65.-k, 42.65.Sf, 52.35.Mw}

\maketitle

\section{introduction}

Nonlinear propagation of light pulses in dielectric media such as
optical fibers has been traditionally modeled using the Nonlinear
Schr\"odinger Equation (NLSE) \cite{agrawal95}. However, it is well
known that NLSE needs modifications to describe a number of higher-order
nonlinear effects which become important at increasing powers and
for short pulses. Recently, the access to new optical systems in which
nonlinearities can be considerably enhanced together with the experimental
availability of ultrashort pulses push the description based on the
NLSE and its modified versions to a limit. A typical example of this
new scenario is provided by the phenomenon of supercontinuum generation
in photonic crystal fibers \cite{ranka-ol25_25}, which requires a
specific modeling that goes beyond approaches based on conventional
versions of the NLSE \cite{husakou-prl87_203901,coen-ol26_1356}.
These new approaches are expressed in new evolution equations that
differ from the NLSE in the amount of approximations needed to achieve
them. We can mention the so-called generalized NLSE \cite{blow-ieeejqe25_2665,mamyshev-ol15_1076},
the nonlinear envelope equation (NEE) \cite{brabec-prl78_3282}, the
forward Maxwell equation (FME) \cite{coen-ol26_1356} and the unidirectional
pulse propagation equation (UPPE) \cite{kolesik-prl89_283902}. Briefly,
the aim of these equations is to describe pulse propagation in the
regime where the frequency width of the pulse is comparable to the
carrier frequency, which, in turn, translates into the fact that usual
approaches suchas the slowly varying approximation no longer hold.
The specific form of these equations is, on the one hand the result
of applying some other different approximations\emph{, e.g,} assuming
propagation in an homogeneous medium (NEE and UPPE) or single-mode
propagation in fibers (generalized NLSE and FME). On the other hand,
two common features of all of them are the neglect of the backward
components of the electromagnetic field and their first-order character. 

The role played by backward components has been previously analized
for an homogeneous medium \cite{shen84}. In this paper we explicitly
unveil their role but in the more general case of an axially-invariant
inhomogenous nonlinear medium by explicitly finding the coupled first-order
equations that drive the forward and backward components of the electromagnetic
field in an axially-invariant nonlinear system. We will show that
these first order forward-backward equations (FBEs) are equivalent
to the original second-order equations for the electric components
of the electromagnetic field. The FBEs will provide us with a common
framework that can encompass different nonlinear phenomena. In fact,
since these equations explicitly manifest the couplings between spatial
and time-frequency degrees of freedom typical of spatial-temporal
phenomena, they can be easily particularized to describe either purely
temporal or purely spatial effects within the same framework, revealing
the total generality of this formalism.

The paper is organized as follows. In section \ref{sec:Modal-second-order-equation}
we derive the most general modal second-order equation for nonlinear
propagation in an axially-invariant inhomogenous medium and we explain
the nature of the only approximation needed to obtain it. In section
\ref{sec:Derivation-of-forward-backward} we demonstrate the equivalence
between the modal second-order equation and the first-order FBEs.
In section \ref{sec:Spinor-representation-of} we introduce a spinor
representation to obtain a Dirac-like form of the FBEs. In section
\ref{sec:Conserved-quantities} we derive the conserved quantities
associated to the FBEs and analize them in the light of phase symmetries.
Finally, in section \ref{sec:Particular-cases} we examine two different
nonlinear phenomena occurring in axially-invariant inhomogeneous dielectric
media in the light of the FBEs: forward pulse propagation and non-paraxial
evolution of spatial structures.

\section{modal second-order equation\label{sec:Modal-second-order-equation}}

The most general equation for the propagation of the electric components
of a electromagnetic field in a inhomogeneous, isotropic and spatially
local dielectric medium is given by\begin{equation}
\nabla^{2}\mathbf{E}_{\omega}-\nabla(\nabla\cdot\mathbf{E}_{\omega})+k_{0}^{2}n_{0}^{2}\mathbf{E}_{\omega}=-\mathbf{G}_{\omega}^{NL}(\mathbf{E}),\label{eq:general_wave_equation}\end{equation}
where $\mathbf{E}_{\omega}=\mathbf{E}_{\omega}(x,y,z)$ is the complex
$\omega$ frequency component of the real electric field $\mathbf{E}(x,y,z,t)=\int_{\omega}\mathbf{E}_{\omega}(x,y,z)\exp(-i\omega t)$,
$\nabla$ is the three-dimensional spatial gradient operator, $k_{0}$
is the vacuum wavenumber, $n_{0}=n_{0}(x,y,z)$ is the refractive
index profile of the dielectric medium and $\mathbf{G}_{\omega}^{NL}$
is the function that represents the local nonlinear response of the
medium to the propagating field. In the most general case, the relative
dielectric constant ($\epsilon=n_{0}^{2}$) and the nonlinear function
$\mathbf{G}_{\omega}^{NL}$ can be either spatially local (the case
we are considering) or non-local if there exist some type of spatially-delayed
response effects. Besides the general wave equation (\ref{eq:general_wave_equation}),
Maxwell's equations require the constraint $\nabla\cdot\mathbf{D}_{\omega}=0$
to be satisfied. Through the constitutive relations, the displacement
field $\mathbf{D}_{\omega}$ is itself a function of the propagating
electric field $\mathbf{E}_{\omega}$. Assuming a spatially local
response, $\mathbf{D}_{\omega}=\epsilon(\mathbf{E}_{\omega})\mathbf{E}_{\omega}$.
In the case the system presented some type of anisotropy, $\epsilon(\mathbf{E}_{\omega})$
would be a second-order tensor. Here we consider an isotropic medium,
so that $\epsilon(\mathbf{E}_{\omega})$ will be a scalar function,
although the generalization to an anisotropic medium is straightforward.
The constraint relation has thus the form:\begin{equation}
\nabla\cdot\left(\epsilon(\mathbf{E}_{\omega})\mathbf{E}_{\omega}\right)=0.\label{eq:constraint}\end{equation}
 In most cases in nonlinear optics, the general constraint (\ref{eq:constraint})
is replaced by the approximated, and much simpler, {}``scalar''
condition $\nabla\cdot\left(\epsilon(\mathbf{E}_{\omega})\mathbf{E}_{\omega}\right)\approx\epsilon(\mathbf{E}_{\omega})\nabla\cdot\mathbf{E}_{\omega}=0$,
implying that $\nabla\cdot\mathbf{E}_{\omega}\approx0$. This approximation
is known to work well in a panoply of nonlinear effects with few and
remarkable exceptions, related mainly to extreme self-focussing events\cite{kolesik-prl89_283902}.
In order to simplify our approach, this approximation will be assumed
throughout, although our main results will remain valid in a slightlier
general context in which the condition $\nabla\cdot\mathbf{E}_{\omega}\approx0$
will not be strictly necessary. 

Our interest lies on describing propagation in an axially-invariant
system, that is, that in which both the refractive index profile $n_{0}(x,y,z)=n_{0}(x,y)$
as well as all other macroscopic nonlinear structural functions, such
as nonlinear susceptibilities, are $z$-independent. In other words,
we will focus on systems for which the axial $z$-dependence is carried
by the propagating field $\mathbf{E}(x,y,z)$ exclusively. Mathematically,
this property is reflected into the fact that there is no explicit
dependence on $z$ in the nolinear function $\mathbf{G}^{NL}$. In
these systems, the wave equation (\ref{eq:general_wave_equation})
and the constraint $\nabla\cdot\mathbf{E}_{\omega}\approx0$ merge
into a single equation:\begin{equation}
\left(\frac{\partial^{2}}{\partial z^{2}}+\nabla_{t}^{2}+k_{0}^{2}n_{0}^{2}(x,y)\right)\mathbf{E}_{\omega}=-\mathbf{G}_{\omega}^{NL}(\mathbf{E}).\label{eq:wave_equation+constraint}\end{equation}
 Where now $\nabla_{t}$ stands for the transverse gradient operator
and the axial second-order derivative has been made explicit. Let
us now perform a modal expansion of the electric field in terms of
the eigenmodes of the linear $z$-independent system, described by
the linear transverse operator $L_{0}\equiv\nabla_{t}^{2}+k_{0}^{2}n_{0}^{2}(x,y)$.
These modes fulfill the linear eigenvalue equation:\begin{equation}
L_{0}\mathbf{\Phi}_{n}^{\omega}(x,y)=\beta_{n}^{2}(\omega)\mathbf{\Phi}_{n}^{\omega}(x,y).\label{eq:lineal_equation}\end{equation}
 Since $L_{0}$ does not mix the spatial (polarization) components
of the electric field, it is proportional to the identity operator
in polarization space. For this reason, its eigenmodes have a three-fold
degeneracy in polarization indices. Therefore, every multiplet of
eigenmodes is constituted by three linearly independent modes, each
one proportional to a three-dimensional vector belonging to a basis
of $\mathbb{R}^{3}$. For simplicity, we can consider this basis to
be the canonical one $\mathbf{\Phi}_{n(\sigma)}^{\omega}(x,y)=\Phi_{n}^{\omega}(x,y)\mathbf{u}_{(\sigma)}$
($\sigma=1,2,3$) where the components of the canonical basis $\{\mathbf{u}_{(1)},\mathbf{u}_{(2)},\mathbf{u}_{(3)}\}$
satisfy $\mathbf{u}_{(\sigma)a}=\delta_{\sigma a}$ ($a=1,2,3$).
If the system is lossless then $n_{0}=n_{0}^{*}$, and $L_{0}$ is
self-adjoint; if it is not, we consider the real part of $n_{0}$
in Eq.(\ref{eq:lineal_equation}) so that $L_{0}$ is again self-adjoint.
In this way, the set of $L_{0}$ eigenmodes $\{\mathbf{\Phi}_{n(\sigma)}^{\omega}\}$
form an orthogonal basis of functions of the transverse coordinates
that can be used to expand the electric field $\mathbf{E}_{\omega}(x,y,z)$
at a given $z$:\[
\mathbf{E}_{\omega}(x,y,z)=\sum_{n,\sigma}c_{n,\sigma}(\omega;z)\Phi_{n}^{\omega}(x,y)\mathbf{u}_{(\sigma)},\]
or, in components,\begin{equation}
E_{\sigma}^{\omega}(x,y,z)=\sum_{n}c_{n\sigma}(w;z)\Phi_{n}^{\omega}(x,y).\label{eq:modal_expansion}\end{equation}

Of course, the modal expansion (\ref{eq:modal_expansion}) has to
be understood as a generalized form of expansion over the entire spectrum
of the $L_{0}$ operator, in which both discrete and continuum parts
of the spectrum can coexist. In practice however, when dealing with
numerical applications, the continuum part is discretized by means
of some convenient election of boundary conditions, so that the discretized
form of the expansion (\ref{eq:modal_expansion}) is, in fact, the
one that is used. Since the election of boundary conditions is an
artifact of the simulation, one has to make sure that physical results
are independent of them.

Now, we substitute the modal expansion (\ref{eq:modal_expansion})
into Eq.(\ref{eq:wave_equation+constraint}), we multiply this equation
by $\Phi_{m}^{\omega*}$, taking into account that $L_{0}\Phi_{n}^{\omega}=\beta_{n}^{2}$$\Phi_{n}^{\omega}$,
and after integrating the result over the entire transverse space
(considering, without loss of generality, that the functions $\Phi_{n}^{\omega}$
are orthonormal: $\int_{\mathbb{R}^{2}}\Phi_{m}^{\omega*}\Phi_{n}^{\omega}=\delta_{nm}$),
we obtain the evolution equation for the modal expansion coefficients
$c$:\begin{equation}
\left(\frac{\partial^{2}}{\partial z^{2}}+\beta_{n}^{2}(\omega)\right)c_{n\sigma}(\omega;z)=-F_{n\sigma}^{\omega}(c).\label{eq:second_order_equation_cont}\end{equation}
 The equation above represents an effective lower-dimensional equation
($1+1$ dimensions, instead of the $3+1$ dimensions of the original
equation (\ref{eq:wave_equation+constraint})) for the coefficients
$c_{n\sigma}(\omega;z)$. Physically, they represent the frequency,
modal and polarization spectral content of the propagating electric
field. The nonlinear function $F_{n\sigma}^{\omega}(c)$ is given
by:\begin{equation}
F_{n\sigma}^{\omega}(c)=\int_{\mathbb{R}^{2}{}}\Phi_{n}^{\omega*}G_{\omega\sigma}^{NL}(\sum c_{n\sigma}\Phi_{n}^{\omega}).\label{eq:definition_of_F}\end{equation}
This function is an input of the effective equation (\ref{eq:second_order_equation_cont})
because it is known once the linear amplitudes are determined out
of the linear eigenvalue problem in Eq.(\ref{eq:lineal_equation}).
Same applies to the propagation constants $\beta_{n}(\omega)$, which
also appear as an input. 

As an example of nonlinear function $F_{n\sigma}^{\omega}(c)$, it
is instructive to consider the case of a Kerr nonlinearity, for which
$G_{\omega\sigma}^{NL}=\omega^{2}/c^{2}\int_{\omega_{2}\omega_{3}}\chi_{\sigma\tau'\sigma'\tau}^{(3)}(\omega,\omega_{2},\omega_{3},\omega+\omega_{2}-\omega_{3})E_{\tau'}^{\omega_{2}*}E_{\sigma'}^{\omega_{3}}E_{\tau}^{\omega+\omega_{2}-\omega_{3}}$.
From the definition of the nonlinear function (\ref{eq:definition_of_F}),
one finds that $F$ has the form:\begin{eqnarray*}
\lefteqn{F_{n\sigma}^{\omega}(c)=\sum_{m'n'm}\,\sum_{\tau'\sigma'\tau}\times}\\
 &  & \!\!\!\!\!\!\!\int_{\omega_{2}\omega_{3}}\!\! c_{m'\tau'}^{\omega_{2}*}\, T_{\stackrel{n\, m'\, n'\, m}{\sigma\,\tau'\,\sigma'\,\tau}}(\omega,\omega_{2},\omega_{3},\omega+\omega_{2}-\omega_{3})\, c_{n'\sigma'}^{\omega_{3}}c_{m\tau}^{\omega+\omega_{2}-\omega_{3}}.\end{eqnarray*}
Thus, for a Kerr nonlinearity, all the information about the nonlinear
properties of the system is encoded in the tensor function $T$. As
a first approximation, one could say that this tensor function depends
on frequency both through the susceptibility $\chi^{(3)}$ and through
the overlapping integral of linear mode amplitudes:\begin{eqnarray*}
\lefteqn{T_{\stackrel{n\, m'\, n'\, m}{\sigma\,\tau'\,\sigma'\,\tau}}(\omega,\omega_{2},\omega_{3},\omega+\omega_{2}-\omega_{3})\approx\omega^{2}/c^{2}}\\
 &  & \!\!\!\!\!\!\!\chi_{\sigma\tau'\sigma'\tau}^{(3)}(\omega,\omega_{2},\omega_{3},\omega+\omega_{2}-\omega_{3})\int_{\mathbb{R}^{2}}\!\!\Phi_{n}^{\omega*}\Phi_{m'}^{\omega_{2}*}\Phi_{n'}^{\omega_{3}}\Phi_{m}^{\omega+\omega_{2}-\omega_{3}}.\end{eqnarray*}
Besides, the dependence on modal indices of the $T$ tensor is a result
of the overlapping of spatial amplitudes, whereas polarization indices
are provided by the third-order susceptibility. 

We do not want to get into much detail on the particular form of the
nonlinear function since, for our purposes, it is not necessary to
provide an explicit form for $F$ in Eq.(\ref{eq:second_order_equation_cont}).
Only general properties of $F$ will be used and the latter can be
inferred from the construction previously described. Along the same
line of reasoning, it should be added that the effective Eq.(\ref{eq:second_order_equation_cont})
remains valid in some cases when the {}``scalar'' approximation
$\nabla\cdot\mathbf{E}_{\omega}\approx0$ no longer holds. Less restrictive
approximations can be made in which $\nabla\cdot\mathbf{D}_{\omega}\approx0$
but $\nabla\cdot\mathbf{E}_{\omega}\neq0$ and, nevertheless, the
effective Eq.(\ref{eq:second_order_equation_cont}) retains, formally,
its validity. It would differ in the fact that the values of $\beta$
would be now calculated including the missing vector term in the linear
eigenvalue equation (\ref{eq:lineal_equation}) and that the nonlinear
function $F$ would include also terms of vectorial origin. Again,
the analysis of such topics is out of the scope of this paper, for
our aim is to transform the almost completely general second-order
equation (\ref{eq:second_order_equation_cont}) into a first-order
formalism in which forward and backward components explicitly appear.

\section{derivation of forward-backward equations\label{sec:Derivation-of-forward-backward}}

A convenient way of considering the effective equation (\ref{eq:second_order_equation_cont})
is as a limit of a equation for a frequency-discretized spectral function
$c_{n\sigma}^{j}(z)\equiv c_{n\sigma}(\omega_{j};z)$. This is a typical
situation in numerical simulations in which frequency appears discretized
in a Fourier series and one approaches the continuum-frequency limit
numerically. Or, likewise in some experimental cases where one physically
works with frequency combs instead of continuum sources. In such situations,
the frequency discretized version of Eq.(\ref{eq:second_order_equation_cont})
is:\begin{equation}
\left(\frac{d^{2}}{dz^{2}}+\beta_{nj}^{2}\right)c_{n\sigma}^{j}(z)=-F_{n\sigma}^{j}(c).\label{eq:second_order_equation}\end{equation}
 The general second-order equation (\ref{eq:second_order_equation}),
displaying all frequency, modal and polarization indices, will be
our starting point. The function $F(c)$ includes all nonlinear contributions
in the frequency-domain constructed out of Maxwell's equations according
to the procedure described in the previous section. As seen above,
for a Kerr nonlinearity in Maxwell's equation, this function would
provide a cubic behavior in the spectral function $c$, which in the
discretized frequency version would be:\begin{eqnarray}
\lefteqn{F_{n\sigma}^{j}(c)=\sum_{m'n'm}\,\sum_{\tau'\sigma'\tau}\,\sum_{k'j'}\times}\nonumber \\
 &  & c_{m'\tau'}^{k'*}\,\left(T_{\stackrel{n\, m'\, n'\, m}{\sigma\,\tau'\,\sigma'\,\tau}}^{jk'l(j+k'-j')}\right)\, c_{n'\sigma'}^{l}c_{m\tau}^{j+k'-j'},\label{eq:example_of_F}\end{eqnarray}
where $T$ is a tensor in frequency, modal and polarization indices
that can be explicitly calculated out of the amplitudes of the linear
fiber modes. However, in order to obtain the first order counterpart
of the general second-order equation (\ref{eq:second_order_equation}),
it is not necessary to specify a particular form for the nonlinearity
function $F$. It is interesting to stress again that both the modal
dispersion relations $\beta_{n}(\omega)$ (or its discretized-frequency
version $\beta_{nj}$) and $F$ are inputs that are known once the
mode linear problem is solved. In order to simplify the notation,
we will incorporate the polarization index into the modal one when
dealing with the spectral function $c$ and the nonlinear function
$F$. This is possible because polarization and modal indices always
come together in pairs. So that, from now on and without loss of generality,
$n$ represents the index pair ($n,\sigma$) . 

We perform now an axial Fourier transform on the spectral functions
$c$ in Eq.(\ref{eq:second_order_equation}), defined as

\[
c_{n}^{j}(z)=\frac{1}{2\pi}\int d\beta\,\tilde{c}_{n}^{j}(\beta)e^{i\beta z},\]

or, inversely,\[
\tilde{c}_{n}^{j}(\beta)=\int dz\, c_{n}^{j}(z)e^{-i\beta z}.\]

With this definition, $\beta\rightarrow-id/dz$ (or, $d/dz\rightarrow i\beta$),
and, thus, the axial Fourier transfom of Eq.(\ref{eq:second_order_equation})
is given by\begin{equation}
(-\beta^{2}+\beta_{nj}^{2})\tilde{c}_{n}^{j}(\beta)=-\tilde{F}_{n}^{j}(\tilde{c}),\label{eq:axial_fourier_equation}\end{equation}
where $\tilde{F}(\tilde{c})$ is the axial Fourier transform of the
nonlinear term in Eq.(\ref{eq:second_order_equation}): $\tilde{F}\equiv\mathcal{F}_{\beta}[F]$.
Dividing this equation by $-\beta^{2}+\beta_{nj}^{2}$(we assume some
kind of {}``$i\epsilon$'' prescription to deal with the poles of
the inverse of this function in the $\beta$ plane), we obtain

\begin{equation}
\tilde{c}_{n}^{j}(\beta)=\frac{1}{\beta^{2}-\beta_{nj}^{2}}\tilde{F}_{n}^{j}(\tilde{c})\label{eq:inverse_eq}\end{equation}

The function preceding the nonlinear factor in the above equation
is the axial Green function in $\beta$-space and has single poles
at $+\beta_{nj}$ and $-\beta_{nj}$, corresponding to forward and
backward propagation, respectively, for the positive frequency part
of the spectral function $\tilde{c}$ (peaked at a positive frequency
$+\omega_{0}$). Clearly, we can decompose the Green function in the
contributions corresponding to the two poles by means of the following
identity:

\begin{equation}
\frac{1}{(\beta^{2}-\beta_{nj}^{2})}=\frac{1}{2\beta_{nj}}(\frac{1}{\beta-\beta_{nj}}-\frac{1}{\beta+\beta_{nj}}),\label{eq:green_function_decomposition}\end{equation}
which allows us to write Eq.(\ref{eq:inverse_eq}) as

\[
\tilde{c}=\tilde{c}_{F}+\tilde{c}_{B},\]

where

\[
\left(\tilde{c}_{F}\right)_{n}^{j}=\frac{1}{2\beta_{nj}}\frac{1}{\beta-\beta_{nj}}\tilde{F}_{n}^{j}(\tilde{c})\]

\[
\left(\tilde{c}_{B}\right)_{n}^{j}=-\frac{1}{2\beta_{nj}}\frac{1}{\beta+\beta_{nj}}\tilde{F}_{n}^{j}(\tilde{c}),\]

or, equivalently,

\[
(\beta-\beta_{nj})\left(\tilde{c}_{F}\right)_{n}^{j}=\frac{1}{2\beta_{nj}}\tilde{F}_{n}^{j}(\tilde{c})\]
\[
(\beta+\beta_{nj})\left(\tilde{c}_{B}\right)_{n}^{j}=-\frac{1}{2\beta_{nj}}\tilde{F}_{n}^{j}(\tilde{c}).\]

Now, we take the inverse axial Fourier transform of the above equations,
taking into account the previous definitions. So that, considering
that $\beta\rightarrow-id/dz$, $\mathcal{F}_{z}^{-1}(\tilde{c}_{F,B})=\mathcal{F}_{z}^{-1}(\mathcal{F}_{\beta}(c_{F,B}))=c_{F,B}$,
and $\mathcal{F}_{z}^{-1}(\tilde{F})=\mathcal{F}_{z}^{-1}(\mathcal{F}_{\beta}(F))=F$,
we can write the following two first-order equations for forward (F)
and backward (B) components: 

\begin{equation}
(-i\frac{d}{dz}-\beta_{nj})\left(c_{F}\right)_{n}^{j}=\frac{1}{2\beta_{nj}}F_{n}^{j}(c)\label{eq:fisrt_order_equation(a)}\end{equation}
\begin{equation}
(-i\frac{d}{dz}+\beta_{nj})\left(c_{B}\right)_{n}^{j}=-\frac{1}{2\beta_{nj}}F_{n}^{j}(c),\label{eq:first_order_equation(b)}\end{equation}
where the total spectrum is given by the sum of forward and backward
contributions

\begin{equation}
c=c_{F}+c_{B}.\label{eq:first_order_equation(c)}\end{equation}

Eqs. (\ref{eq:fisrt_order_equation(a)}) and (\ref{eq:first_order_equation(b)})
are the exact first-order equations equivalent of the general second-order
equation (\ref{eq:second_order_equation}).

Restoring the continuum frequency notation ($c^{j}\rightarrow c(\omega)$),
we obtain the first-order forward-backward equations in the continuum-frequency
limit:\begin{equation}
\left(-i\frac{\partial}{\partial z}-\beta_{n}(\omega)\right)c_{F}^{n}(\omega,z)=\frac{1}{2\beta_{n}(\omega)}F_{n}(c(\omega,z))\label{eq:first_order_eq_cont(a)}\end{equation}

\begin{equation}
\left(-i\frac{\partial}{\partial z}+\beta_{n}(\omega)\right)c_{B}^{n}(\omega,z)=-\frac{1}{2\beta_{n}(\omega)}F_{n}(c(\omega,z))\label{eq:first_order_eq_cont(b)}\end{equation}
whereas the total spectrum is\[
c(\omega,z)=c_{F}(\omega,z)+c_{B}(\omega,z).\]

Notice that if the backward spectrum is sufficiently small and it
can be approximately neglected ($c_{B}\approx0)$, then $c\approx c_{F}$
and only Eqs. (\ref{eq:fisrt_order_equation(a)}) or (\ref{eq:first_order_eq_cont(a)})
matter. It is also interesting to notice that small values of the
backward spectrum imply small nonlinearities, as it can be checked
by going to the $c_{B}\rightarrow0$ limit in Eqs.(\ref{eq:first_order_equation(b)})
or (\ref{eq:first_order_eq_cont(b)}). In such a limit, nonlinearities
dissapear independently of the nature of their origin: $F(c)\rightarrow0$.
We will return to this point in the last section of this paper.

\section{spinor representation of the forward-backward equations \label{sec:Spinor-representation-of}}

It is possible to put the first order FBEs in a more compact form,
by using a spinor representation for the forward and backward spectral
functions, $c_{F}$ and $c_{B}$. We start noticing that the FBEs
(with discrete-frequency indices) can be rewritten as, 

\[
(-i\frac{d}{dz}-\beta_{nj})c_{F}^{nj}=\sum_{n'j'}\left[N_{nn'}^{jj'}(c)c_{F}^{n'j'}+N_{nn'}^{jj'}(c)c_{B}^{n'j'}\right]\]

\begin{equation}
(-i\frac{d}{dz}+\beta_{nj})c_{B}^{nj}=\sum_{n'j'}\left[-N_{nn'}^{jj'}(c)c_{F}^{n'j'}-N_{nn'}^{jj'}(c)c_{B}^{n'j'}\right].\label{eq:first_order_eq_N}\end{equation}

We have assumed that the nonlinear function $F$ can be expressed
as the action of a field-dependent operator $M(c)$ on the spectral
function $c$; that is, $F_{n}^{j}(c)\equiv\sum_{n'j'}M_{nn'}^{jj'}(c)c_{n'}^{j'}$.
This is the situation that applies to all type of nonlinearities that
can be expanded in a power series. The simplest case is the aforementioned
Kerr cubic nonlinearity, for which $M_{nn'}^{jj'}(c)=\sum_{mm'}^{k'}c_{m'}^{k'*}\, T_{nm'n'm}^{jk'j'(j+k'-j')}\, c_{m}^{j+k'-j'}$.
Moreover, in conservative systems, that is, those for which the Hamiltonian
is conserved and real, the $M$ operator is forced to be self-adjoint.The
$N$ operator appearing in Eq.(\ref{eq:first_order_eq_N}) is nothing
but $N_{nn'}^{jj'}(c)\equiv1/(2\beta_{nj})M_{nn'}^{jj'}(c)$. 

The equations above can also be written in a matrix form,

\begin{eqnarray}
\lefteqn{-i\frac{d}{dz}\left(\begin{array}{c}
c_{F}^{nj}\\
c_{B}^{nj}\end{array}\right)-\beta_{nj}\left(\begin{array}{cc}
1 & 0\\
0 & -1\end{array}\right)\left(\begin{array}{c}
c_{F}^{nj}\\
c_{B}^{nj}\end{array}\right)}\nonumber \\
 &  & =\sum_{n'j'}N_{nn'}^{jj'}(c)\left(\begin{array}{cc}
1 & 1\\
-1 & -1\end{array}\right)\left(\begin{array}{c}
c_{F}^{n'j'}\\
c_{B}^{n'j'}\end{array}\right)\label{eq:matrix_equation}\end{eqnarray}

Then, introducing the bi-spinor $\psi$, defined as\begin{equation}
\psi_{n}^{j}\equiv\left(\begin{array}{c}
c_{F}^{nj}\\
c_{B}^{nj}\end{array}\right),\label{eq:spinor}\end{equation}
and the Pauli matrices,

\[
\sigma_{1}=\left(\begin{array}{cc}
0 & 1\\
1 & 0\end{array}\right),\,\,\sigma_{2}=\left(\begin{array}{cc}
0 & -i\\
i & 0\end{array}\right),\,\,\sigma_{3}=\left(\begin{array}{cc}
1 & 0\\
0 & -1\end{array}\right).\]

noticing that

\[
\left(\begin{array}{cc}
1 & 1\\
-1 & -1\end{array}\right)=\sigma_{3}+i\sigma_{2},\]
we see that Eq.(\ref{eq:matrix_equation}) can be written in a bi-spinor
form notation (a sum over repeated indices is assumed)

\begin{equation}
-i\frac{d}{dz}\psi_{n}^{j}=\left[\left(\beta_{n'j'}\delta_{nn'}\delta_{jj'}+N_{nn'}^{jj'}\right)\sigma_{3}+iN_{nn'}^{jj'}\sigma_{2}\right]\psi_{n'}^{j'}.\label{eq:first_order_equation_spinor}\end{equation}

Certainly, $N$ is a non-trivial operator in frequency and mode spaces.
However, it is proportional to the identity operator when acting on
the $F-B$ internal degrees of freedom of the bi-spinor $\psi$. If
we use continuum-frequency notation, Eq.(\ref{eq:first_order_equation_spinor})
can be written as\begin{eqnarray}
\lefteqn{-i\frac{\partial}{\partial z}\psi_{n}(\omega,z)=\beta_{n}(\omega)\sigma_{3}\psi_{n}(\omega,z)+}\nonumber \\
 &  & \sum_{n'}\int d\omega'N_{nn'}(\omega,\omega';\psi)(\sigma_{3}+i\sigma_{2})\psi_{n'}(\omega',z).\label{eq:first_order_equation_spinor_cont}\end{eqnarray}
Despite its different appearence, the first-order spinor equation
for the pulse spectrum contains exactly the same information on dynamics
than the original second-order equation (\ref{eq:second_order_equation}).
The spinor representation of the FBEs has common features with the
Dirac equation for a particle in $1+1$ dimensions. In $1+1$ dimensions,
the original algebra of 16 $4\times4$ matrices (Dirac matrices) of
the 4D Dirac equation is reduced to 4 $2\times2$ matrices constituted
by the identity and Pauli matrices. The FBEs spinor (\ref{eq:spinor}),
formed by the forward and backward spectral functions, plays the role
of the Dirac spinor, constituted by the positive and negative energy
components of the particle wave-function.

\section{conserved quantities \label{sec:Conserved-quantities}}

The goal of this section is to find the conserved quantity associated
to FBEs in the most general case. Our only assumption will be the
conservative and lossless character of the system, which mathematically
will be reflected in the fact that the $M$ operator is self-adjoint
($M=M^{\dagger}$) and that the dispersion relation function $\beta$
is real for all modes. We consider as a starting point the discrete-frequency
FBEs in their spinor representation (\ref{eq:first_order_equation_spinor}).
Now we proceed to redefine the labeling of this equation following
the same procedure we used to incorporate polarization indices into
modal indices in Section \ref{sec:Derivation-of-forward-backward}.
Since frequency ($j$) and modal indices ($n$) are always paired
together in Eq.(\ref{eq:first_order_equation_spinor}), we can incorporate
both in a new single index: $(n,j)\rightarrow p$, $(n',j')\rightarrow p$'.
In this way, Eq.(\ref{eq:first_order_equation_spinor}) becomes (a
sum over repeated indices is assumed, as before):\begin{equation}
-i\frac{d}{dz}\psi_{p}=\left[\left(\beta_{p'}\delta_{pp'}+N_{pp'}\right)\sigma_{3}+iN_{pp'}\sigma_{2}\right]\psi_{p'}.\label{eq:spinor_equation_new_label}\end{equation}
This form of the equation enables us to introduce the matrix operators
$\Psi$, B and N, whose elements are given by $\psi_{p}$, $\beta_{p'}\delta_{pp'}$
and $N_{pp'}$, respectively. Therefore, we can write ($\dot{\Psi}\equiv d\Psi/dz$):\begin{equation}
-i\dot{\Psi}=\left\{ \left(\mathrm{B+\frac{1}{2}B^{-1}M}\right)\sigma_{3}+i\mathrm{\frac{1}{2}B^{-1}M}\sigma_{2}\right\} \Psi,\label{eq:spinor_equation_matrix_form}\end{equation}
where we have made use of the relation between the $N$ and $M$ operators:
$N_{pp'}=1/(2\beta_{p})M_{pp'}\Rightarrow\mathrm{N=(1/2)B^{-1}M}$.
We will also need the adjoint of the above equation:\begin{equation}
i\dot{\Psi}^{\dagger}=\Psi^{\dagger}\left\{ \left(\mathrm{B+\frac{1}{2}MB^{-1}}\right)\sigma_{3}-i\mathrm{\frac{1}{2}MB^{-1}}\sigma_{2}\right\} ,\label{eq:spinor_equation_matrix_form_adjoint}\end{equation}
where we have used the properties that $\mathrm{B=B^{\dagger}}$(since
B is diagonal and $\beta$ is real), $\mathrm{M=M^{\dagger}}$, together
with the self-adjointness of the Pauli matrices. Notice that both
B and M are proportional to the identity matrix when they act on the
$F-B$ components of the bi-spinor $\Psi$, and for this reason they
commute with Pauli matrices. Next, we left-multiply Eq.(\ref{eq:spinor_equation_matrix_form})
by $\Psi^{\dagger}\mathrm{B}\sigma_{3}$ and right-multiply Eq.(\ref{eq:spinor_equation_matrix_form_adjoint})
by $\mathrm{B}\sigma_{3}\Psi$, to obtain ($\sigma_{3}^{2}=1$, $\sigma_{2}\sigma_{3}=i\sigma_{1}$):\[
i\dot{\Psi}^{\dagger}\mathrm{B}\sigma_{3}\Psi=\Psi^{\dagger}\left\{ \mathrm{B^{2}}+\frac{1}{2}\mathrm{M}+\frac{1}{2}\mathrm{M}\sigma_{1}\right\} \Psi\]
\[
-i\Psi^{\dagger}\mathrm{B}\sigma_{3}\dot{\Psi}=\Psi^{\dagger}\left\{ \mathrm{B^{2}}+\frac{1}{2}\mathrm{M}+\frac{1}{2}\mathrm{M}\sigma_{1}\right\} \Psi\]

We achieve the desired result by substracting the previous equations:\begin{equation}
i\left(\dot{\Psi}^{\dagger}\mathrm{B}\sigma_{3}\Psi+\Psi^{\dagger}\mathrm{B}\sigma_{3}\dot{\Psi}\right)=0\Leftrightarrow\frac{d}{dz}\left(\Psi^{\dagger}\mathrm{B}\sigma_{3}\Psi\right)=0.\label{eq:conservation_law}\end{equation}

The conserved quantity is thus $Q=\Psi^{\dagger}\mathrm{B}\sigma_{3}\Psi$,
or, after reintroducing indices, $Q=\sum_{n,j}\psi_{nj}^{\dagger}\beta_{nj}\sigma_{3}\psi_{nj}$
(discrete frequency) or $Q=\sum_{n}\int d\omega\psi_{n}^{\dagger}(\omega)\beta_{n}(\omega)\sigma_{3}\psi_{n}(\omega)$
(continuous frequency). In terms of the original forward and backward
components of the bi-spinor $\psi$, the conserved quantity has the
following form:\begin{eqnarray}
\lefteqn{Q} &  & =\sum_{n}\int d\omega\times\label{eq:Q}\\
 &  & \beta_{n}(\omega)\left(c_{Fn}^{*}(\omega,z)c_{Fn}(\omega,z)-c_{Bn}^{*}(\omega,z)c_{Bn}(\omega,z)\right).\nonumber \end{eqnarray}

The physical meaning of this conserved quantity can help us to understand
its particular form. Eq.(\ref{eq:Q}) is the modal frequency version
of the axial component of the Poynting vector, as one could check
by reminding that, for the case of Maxwell's equations in the scalar
approximation, $\mathcal{P}_{z}\sim\int_{\mathbb{R}^{2}}i\phi^{*}\partial_{z}\phi$.
This represents the amount of electromagnetic energy traversing a
section of the system per unit time. For this reason, since we are
dealing in fact with the axial flux of the electromagnetic field,
it is natural that the quantity $Q$, the total electromagnetic axial
flux, can be considered as the sum of the positive forward axial flux
($+Q_{F}$) and the negative backward axial flux ($-Q_{B}$): $Q=Q_{F}-Q_{B}$,
where $Q_{F}=\sum_{n}\int\beta_{n}c_{Fn}^{*}c_{Fn}$ and $Q_{B}=\sum_{n}\int\beta_{n}c_{Bn}^{*}c_{Bn}$.

\section{particular cases \label{sec:Particular-cases}}

An interesting case to consider is that occuring when one neglects
all $F-B$ interactions. In general, the FBEs (\ref{eq:first_order_eq_cont(a)})
and (\ref{eq:first_order_eq_cont(b)}) mix $F-B$ components due to
the presence of the nonlinear function $F(c)=F(c_{F}+c_{B})$ in the
right hand side of both equations. When $F-B$ interactions can be
neglected, what happens when either $c_{F}$ or $c_{B}$ are very
small, FBEs decouple and we obtain two separate equations for $c_{F}$
and $c_{B}$. In such a case, forward and backward axial fluxes are
conserved independently: $dQ_{F}/dz=0$ and $dQ_{B}/dz=0$. The demonstration
of this property closely follows the general proof previously described
for the total flux $Q$. Let us consider FBEs in their discrete-frequency
form (Eqs.(\ref{eq:fisrt_order_equation(a)}) and (\ref{eq:first_order_equation(b)}))
when all $F-B$ mixing terms are neglected; that is, when $F(c)\approx F(c_{F})$
in Eq.(\ref{eq:fisrt_order_equation(a)}) because $c\approx c_{F}$
($c_{B}\approx0$) or $F(c)\approx F(c_{B})$ in Eq.(\ref{eq:first_order_equation(b)})
because $c\approx c_{B}$ ($c_{F}\approx0$). We will consider forward
components only (backward analysis is completely analogous). By relabelling
the frequency and modal indices together into a new index and by introducing
the matrix notation in the same way we used before to obtain Eq.(\ref{eq:spinor_equation_matrix_form}),
the equation for forward components adopts the form:\begin{equation}
-i\dot{\mathrm{c}}_{\mathrm{F}}=\left(\mathrm{B}+\frac{1}{2}\mathrm{B^{-1}M(c_{F})}\right)\mathrm{c_{F}}.\label{eq:equation_forward_matrix}\end{equation}
In this case, the equation for the backward component would correspond
to a very weak field ($c_{B}\approx0$) and, thus, it would be basically
linear: $-i\dot{\mathrm{c}}_{\mathrm{B}}\approx-\mathrm{Bc_{B}}$.
For our purposes, we also need the adjoint equation of (\ref{eq:equation_forward_matrix}):
\[
i\dot{\mathrm{c}}_{\mathrm{F}}^{\dagger}=\mathrm{c_{F}}^{\dagger}\left(\mathrm{B}+\frac{1}{2}\mathrm{M(c_{F})B^{-1}}\right).\]
Left-multiplying Eq.(\ref{eq:equation_forward_matrix}) by $\mathrm{c_{F}}^{\dagger}\mathrm{B}$
and right-multiplying the previous equation by $\mathrm{Bc_{F}}$,
one gets:\[
i\mathrm{c_{F}}^{\dagger}\mathrm{B}\dot{\mathrm{c}}_{\mathrm{F}}=\mathrm{c_{F}}^{\dagger}\left(\mathrm{B^{2}}+\frac{1}{2}\mathrm{M(c_{F})}\right)\mathrm{c_{F}}\]
\[
-i\mathrm{\dot{c}_{F}}^{\dagger}\mathrm{B}\mathrm{c}_{\mathrm{F}}=\mathrm{c_{F}}^{\dagger}\left(\mathrm{B^{2}}+\frac{1}{2}\mathrm{M(c_{F})}\right)\mathrm{c_{F}}.\]
The desired conservation law comes now after substracting the above
equations:\[
\frac{d}{dz}(\mathrm{c_{F}^{\dagger}Bc_{F}})=\mathrm{c_{F}}^{\dagger}\mathrm{B}\dot{\mathrm{c}}_{\mathrm{F}}+\mathrm{\dot{c}_{F}}^{\dagger}\mathrm{B}\mathrm{c}_{\mathrm{F}}=0.\]
 Certainly, $Q_{F}=\mathrm{c_{F}^{\dagger}Bc_{F}}=\sum_{n}\int\beta_{n}c_{Fn}^{*}c_{Fn}$
is the conserved forward axial flux. Conservation of $Q_{B}=\mathrm{c_{B}^{\dagger}Bc_{B}}=\sum_{n}\int\beta_{n}c_{Bn}^{*}c_{Bn}$
is a trivial issue because $\mathrm{c_{B}}$ fulfills the linear equation
$-i\dot{\mathrm{c}}_{\mathrm{B}}\approx-\mathrm{Bc_{B}}$. An analogous
proof shows that $Q_{F}$ and $Q_{B}$ are also conserved when forward
components are neglected instead. 

An alternative analysis can be formulated in the light of conserved
phase symmetries when $F-B$ interactions are neglected. Both forward
and backward axial fluxes can be also envisaged as the conserved charges
associated to independent global $U(1)$ symmetry transformations
on $F$ and $B$ components, respectively. This is clear in Eq.(\ref{eq:equation_forward_matrix}),
which is invariant under $U(1)_{F}$ global phase transformations

\begin{equation}
\mathrm{c_{F}}\rightarrow e^{i\theta_{\mathrm{F}}}\mathrm{c_{F}}\label{eq:phase_transf}\end{equation}
when $\mathrm{M(c_{F})=M(e^{i\theta_{F}}c_{F})}$. This is the situation
for all type of nonlinearities that can be expanded in odd power series
in conservative lossless systems. Again, the simplest example is a
cubic or Kerr nonlinearity, for which $\mathrm{M(c)=c^{*}Tc}$. Since
the equation for backward components is basically linear ($-i\dot{\mathrm{c}}_{\mathrm{B}}\approx\mathrm{Bc_{B}}$),
global $U(1)_{B}$ invariance ($\mathrm{c_{B}}\rightarrow e^{i\theta_{\mathrm{B}}}\mathrm{c_{B}}$)
is trivially fulfilled as well. $U(1)_{F}\otimes U(1)_{B}$ invariance
requires the independent conservation of its associated $U(1)_{F}$
and $U(1)_{B}$ charges, which are nothing but the $F$ axial flux
($Q_{F}$) and the $B$ axial flux ($-Q_{B}$); that is, $dQ_{F}/dz=0$
and $-dQ_{B}/dz=0$. If the situation were the opposite and forward
components were neglected, a complete analogous analysis for backward
components would hold leading to the same conclusion.

When $F-B$ interactions cannot be neglected, the previous argument
does not hold and neither $Q_{F}$ nor $Q_{B}$ are conserved separately
in conservative lossless systems. From a symmetry point of view, this
can be understood by the fact that FBEs (\ref{eq:first_order_eq_cont(a)})
and (\ref{eq:first_order_eq_cont(b)}) are no longer invariant under
independent $F-B$ phase transformations ($c=c_{F}+c_{B}\rightarrow\tilde{c}=e^{i\theta_{F}}c_{F}+e^{i\theta_{B}}c_{B}$)
because in that case $M(\tilde{c})=M(e^{i\theta_{F}}c_{F}+e^{i\theta_{B}}c_{B})\neq M(c_{F}+c_{B})=M(c)$,
as can be easily checked for a cubic nonlinearity. Using the language
of group theory, one would state that $U(1)_{F}\otimes U(1)_{B}$
symmetry is broken. However, there is a residual phase symmetry remaining
in FBEs (\ref{eq:first_order_eq_cont(a)}) and (\ref{eq:first_order_eq_cont(b)})
when $U(1)_{F}\otimes U(1)_{B}$ invariance is broken by $F-B$ interactions.
FBEs are still invariant under simultaneous global phase tranformations
on $F$ and $B$ components ($c_{F}\rightarrow e^{i\theta}c_{F}$
and $c_{B}\rightarrow e^{i\theta}c_{B}$, which implies $c\rightarrow e^{i\theta}c$)
provided that $M(e^{i\theta}c)=M(c)$. Notice that this $U(1)$ symmetry
is a particular case of the higher-order $U(1)_{F}\otimes U(1)_{B}$
symmetry when $\theta_{F}=\theta$ and $\theta_{B}=\theta$. Since
$U(1)_{F}\otimes U(1)_{B}$ symmetry is broken, the $U(1)_{F}$ charge,
i.e, the $F$ axial flux ($Q_{F}$), and the $U(1)_{B}$ charge, i.e.,
the $B$ axial flux ($-Q_{B}$), are not conserved. However, in such
a situation the remaining $U(1)$ symmetry guarantees the conservation
of a new quantity, namely, the \emph{sum} of the $U(1)_{F}$ and $U(1)_{B}$
charges. That is, $Q_{F}+(-Q_{B})$ has to be conserved. The total
axial electromagnetic flux $Q=Q_{F}-Q_{B}$ appears then as the conserved
charge of the $U(1)$ symmetry associated to the breaking pattern
$U(1)_{F}\otimes U(1)_{B}\rightarrow U(1)_{F+B}$.

There are two different and interesting physical situations in which
the analysis described above for systems in which $F-B$ interactions
can be neglected is valid. The first one is general forward pulse
propagation, that naturally takes place in the frequency domain or,
equivalently, in the time domain. The second one is monochromatic
(or quasi-monochromatic) non-paraxial forward evolution describing
the nonlinear propagation of spatial structures. It is remarkable
that despite the distinct nature of both phenomena, they are just
particular cases of FBEs and, thus, equally described by the same
formalism. We will study them separately in the next two subsections.

\subsection{Forward pulse propagation}

When one neglects $F-B$ interactions, for example by eliminating
backward components, the most general form of FBEs is given by Eq.(\ref{eq:equation_forward_matrix}),
that in continuum frequency-notation reads:\begin{eqnarray}
\lefteqn{\lefteqn{}-i\frac{\partial}{\partial z}c_{F}^{n}(\omega,z)=\beta_{n}(\omega)c_{F}^{n}(\omega,z)}\nonumber \\
 &  & +\frac{1}{2\beta_{n}(\omega)}\sum_{n'}\int d\omega'M_{nn'}(\omega,\omega';c_{F})c_{F}^{n'}(\omega',z).\label{eq:forward_pulse_equation}\end{eqnarray}
The nature of the nonlinearities define the particular form of the
nonlinear modal matrix function $M_{nn'}(\omega,\omega')$. According
to what was explained in Section \ref{sec:Derivation-of-forward-backward},
the functional form of the nonlinear function $F$ and, thus, of the
$M$ matrix function, can be systematically constructed out of the
mode amplitudes of the linear propagation problem together with the
standard nonlinear coefficients ($\chi^{(3)},\chi^{(5)}$, $\ldots$).
Thus, many formal properties of $M$ will be inherited from linear
modes. The dependence of $M$ on modal indices and frequency will
have much to do with the particular dependence of the linear mode
amplitudes on spatial coordinates and frequency. Different physically
well-grounded assumptions on the properties of $M$ on modal (and
polarization) indices and frequency can be then made by analyzing
linear mode characteristics. The second element to take into account
is the extension of the frequency spectrum $c_{F}(\omega)$. A considerable
simplification is achieved for sufficiently narrow spectra, whereas
wide bandwiths, as those naturally appearing in highly-nonlinear fibers
(e.g., in supercontinuum generation), would demand to consider the
frequency dependence of Eq.(\ref{eq:forward_pulse_equation}) in its
total extent. 

It is clear that, in the most general case, Eq.(\ref{eq:forward_pulse_equation})
involves an intrincate dynamics since the nonlinear matrix function
$M_{nn'}(\omega,\omega'$) leads to nonzero couplings between different
frequency, modal and even polarization components (recall that polarization
indices are included). Thus, even starting from a simple spectral,
single-mode, single-polarization configuration, if the system had
no physical mechanism to minimize the great variety of couplings induced
by $M$, spectral evolution, as described by Eq.(\ref{eq:forward_pulse_equation}),
would generate a more and more complicated spectral function by exciting
new frequency, modal and polarization components.

An important feature of Eq.(\ref{eq:forward_pulse_equation}) is the
existence of an inherent spatial modal interference of nonlinear origin,
represented by the nondiagonal nature of the nonlinear matrix function
in the modal indices --$M_{nn'}\neq0$ ($n\neq n'$)-- in the most
general case. In some special situations, however, this modal interplay
can be zero or it can be just neglected ($M_{nn'}=0$ or $M_{nn'}\approx0$,
when $n\neq n'$). Exact cancellation of matrix elements occurs by
symmetry considerations in most cases. A simple example would be given
by a rotationally invariant system. Its linear modes are solutions
with well-defined angular momentum and, consequently, the modal index
of the spectral function $c_{F}^{n}$ is thus labelled by the angular
momentum index $l$. In such a case, angular momentum conservation
requires that no mixing of spectral components with different angular
momentum occurs, thus eliminating nondiagonal terms in the nonlinear
matrix function corresponding to different values of the angular momentum
index $l$; that is, $M$ has to be diagonal in these indices: $M_{nn'}\sim\delta_{ll'}$.
In other cases, some nondiagonal matrix elements can be negligible
because of the different shapes of linear modes amplitudes involved
in the calculation of the overlapping integrals appearing in the definition
of $M$ (for example, for a Kerr cubic nonlinearity: $M_{nn'}\sim c_{m'}^{*}c_{m}\int_{R^{2}}\phi_{m'}^{*}\phi_{n'}^{*}\phi_{m}\phi_{n}$).
In some cases, these integrals can be very small for reasons that
do not rely on the presence of particular symmetries.

Whatever these reasons may be, in the case that modal interplay does
not exist, or this can be neglected --that is, when $M_{nn'}=0$ or
$M_{nn'}\approx0$ if $n\neq n'$--, then Eq.(\ref{eq:forward_pulse_equation})
decouples into independent equations for every mode index, \begin{eqnarray}
\lefteqn{\lefteqn{}-i\frac{\partial}{\partial z}c_{F}^{n}(\omega,z)=\beta_{n}(\omega)c_{F}^{n}(\omega,z)}\nonumber \\
 &  & +\frac{1}{2\beta_{n}(\omega)}\int d\omega'M_{n}(\omega,\omega';c_{F})c_{F}^{n}(\omega',z),\label{eq:forward_pulse_equation_n}\end{eqnarray}
which, in turns, leads to independent conservation laws for the different
modal $F$ and $B$ axial fluxes: $dQ_{F}^{(n)}/dz=0$ and $dQ_{B}^{(n)}/dz=0$
($Q_{F}^{(n)}\equiv\int_{\omega}\beta_{n}c_{F}^{(n)*}c_{F}^{(n)}$
and $Q_{B}^{(n)}\equiv\int_{\omega}\beta_{n}c_{B}^{(n)*}c_{B}^{(n)}$). 

In nonlinear propagation in optical single-mode fibers it is commonly
assumed that the spatial dependence of the electric propagating field
is just given by the amplitude of the fundamental mode of the fiber.
In our context, this statement is equivalent to say that the matrix
$M$ is one-dimensional and involves the fundamental mode only. The
equation describing forward nonlinear propagation in such a fiber
would be Eq.(\ref{eq:forward_pulse_equation_n}) for $n=0$ (fundamental
mode). Even if the fiber is not single-mode but involves modes widely
separated by large gaps in $\beta$'s, as in some highly-nonlinear
microestructured fibers, the same equation can still remain approximately
valid. The reason is that in such a case intermodal interactions are
relatively suppressed with respect to modal self-interactions because
of the very different forms of linear mode amplitudes, originated
by both the discrete symmetry of the fiber and their very different
values of $\beta$, leading to zero or small overlapping integrals
and, thus, to small values of $M_{nn'}$ when $n\neq n'$. The resulting
approximated equation ---Eq.(\ref{eq:forward_pulse_equation_n}) restricted
to the fundamental mode ($n=0$)--- is equivalent to the Forward Maxwell
Equation (FME) for a highly-nonlinear microstructured fiber used to
adequately describe supercontinuum generation in this type of fibers
\cite{husakou-prl87_203901}.

\subsection{Non-paraxial spatial evolution}

Monochromatic (or, in practice, quasi-monochromatic) propagation is
also a particular situation that can be described using FBEs. Eqs.
(\ref{eq:first_order_eq_cont(a)}) and (\ref{eq:first_order_eq_cont(b)})
remain certainly valid when the frequency content of the $F$ and
$B$ spectral functions $c_{F}$ and $c_{B}$ is restricted to a single
frequency $\omega_{0}$. In such a case, only intermodal interaction
or modal self-interaction (including polarization) play a role since
the nonlinear matrix function has a trivial dependence on frequency
$M_{nn'}(\omega,\omega';c)=M_{nn'}(\omega_{0};c)$ fixed by the propagation
frequency $\omega_{0}$. Mathematically, FBEs become (in order to
symplify the notation: $\beta_{0n}\equiv\beta_{n}(\omega_{0})$,$M_{nn'}(c)\equiv M_{nn'}(\omega_{0};c)$,
$c(z)\equiv c(z;\omega_{0})$)\begin{eqnarray}
\lefteqn{\lefteqn{}-i\frac{\partial}{\partial z}c_{F}^{n}(z)=\beta_{0n}c_{F}^{n}(z)}\nonumber \\
 &  & +\frac{1}{2\beta_{0n}}\sum_{n'}M_{nn'}(c)c_{n'}(z)\label{eq:forward_equation_fixed_freq}\end{eqnarray}
\begin{eqnarray}
\lefteqn{\lefteqn{}-i\frac{\partial}{\partial z}c_{B}^{n}(z)=-\beta_{0n}c_{B}^{n}(z)}\nonumber \\
 &  & -\frac{1}{2\beta_{0n}}\sum_{n'}M_{nn'}(c)c_{n'}(z),\label{eq:backward_equation_fixed_freq}\end{eqnarray}
and $c=c_{F}+c_{B}$. 

Since the previous FBEs involve spatial modal indices only, it is
interesting to write them also in the spatial domain, so that spatial
degrees of freedom appear explicitly. This process is the inverse
of the one we followed to obtain the second-order modal equation in
Section \ref{sec:Modal-second-order-equation}; that is, we reintroduce
the spatial field amplitudes as $\phi_{F,B}(x,y,z)=\sum_{n}c_{F,B}^{n}(z)\phi_{n}(x,y)$,
$\phi_{n}$ being the eigenfunctions of the linear operator $L_{0}\equiv\ \nabla_{t}^{2}+k_{0}^{2}n_{0}^{2}$
and $\beta_{n0}$ their corresponding eigenvalues evaluated at the
fixed frequency $\omega_{0}=k_{0}/c$. The outcome is\begin{equation}
\left(-i\frac{\partial}{\partial z}-L_{0}^{1/2}\right)\phi_{F}=\frac{1}{2}L_{0}^{-1/2}M(\phi)\phi\label{eq:forward_equation_spatial}\end{equation}
\begin{equation}
\left(-i\frac{\partial}{\partial z}+L_{0}^{1/2}\right)\phi_{B}=-\frac{1}{2}L_{0}^{-1/2}M(\phi)\phi,\label{eq:backward_equation_spatial}\end{equation}
together with $\phi=\phi_{F}+\phi_{B}$. The nonlinear term $M(\phi)\phi$
includes all type of nonlinearities in the spatial domain that permit
an expansion in power series. The usual example is the Kerr nonlinearity
$M(\phi)\phi\sim(\phi^{*}\phi)\phi$. 

It is also an interesting exercise to prove that one can derive the
standard second-order wave equation from the first-order spatial FBEs
(\ref{eq:forward_equation_spatial}) and (\ref{eq:backward_equation_spatial}).
If we sum and substract Eq.(\ref{eq:forward_equation_spatial}) and
Eq.(\ref{eq:backward_equation_spatial}), we obtain\begin{equation}
-i\frac{\partial}{\partial z}(\phi_{F}+\phi_{B})-L_{0}^{1/2}(\phi_{F}-\phi_{B})=0\label{eq:sum_equation}\end{equation}
and\begin{equation}
-i\frac{\partial}{\partial z}(\phi_{F}-\phi_{B})-L_{0}^{1/2}(\phi_{F}+\phi_{B})=L_{0}^{-1/2}M(\phi)\phi,\label{eq:difference_equation}\end{equation}
respectively. By applying ($i\partial/\partial z$) onto Eq.(\ref{eq:sum_equation}),
susbtituting Eq.(\ref{eq:difference_equation}) into the resulting
equation, and taking into account that $\phi=\phi_{F}+\phi_{B}$,
one gets the Helmholtz-type nonlinear wave equation:\[
\left(\frac{\partial^{2}}{\partial z^{2}}+\nabla_{t}^{2}+k_{0}^{2}n_{0}^{2}+M(\phi)\right)\phi=0.\]

As mentioned in Section \ref{sec:Derivation-of-forward-backward},
and it is also evident in Eq.(\ref{eq:backward_equation_spatial}),
the complete elimination of backward components ($\phi_{B}\rightarrow0$)
implies the vanishing of nonlinearities ($M\rightarrow0$). Conversely,
if $M\neq0$ then backward components are necessarily generated: $\phi_{B}\neq0$
(even if they did not initially exist). Therefore, there exists a
close link between backward amplitudes and the nonlinear function
$M$. Certainly, in most experimental situations in which an axially-invariant
system is axially illuminated along a privileged (say, forward) direction
exclusively, backward components are small as compared to forward
amplitudes. However, as backward FBE show, they cannot be identically
zero in the presence of nonlinearities. They would be exactly zero
only in the case the system behaved linearly. In such a case, since
the system is axially invariant and it is illuminated in the forward
direction only, there would be no axial inhomogeneities that could
produce linear reflections. Thus, small backward amplitudes are expected
to be generated by small nonlinearities so that $\phi_{B}\rightarrow0$
when $M\rightarrow0$ (linear case). Our interest lies now in the
quantification of the relation between $\phi_{B}$ and $M$ in the
small backward amplitud regime ($\phi_{B}=\delta\phi_{B}\ll1$).

In order to clarify the calculation, we parametrize the {}``size''
of the nonlinearity by redefining the nonlinear function as $M=\gamma\bar{M}$,
where $\gamma$ is a dimensionless real parameter. In this way, we
can approach the linear case ($M\rightarrow0$) by taking the limit
$\gamma\rightarrow0$. It is easy to realize that $\gamma$ has to
be proportional to the input power $\gamma\sim P$ when $\gamma$
is small. Certainly, $M$ is a function of the input power verifying
that $M\rightarrow0$ as $P\rightarrow0$. Assuming, as usual, that
this dependence is analytical this implies that $M\sim P$ when $P\rightarrow0$.
The same argument for $\gamma$ leads to $M\sim\gamma$ for small
values of $\gamma$. Therefore, the small $\gamma$ and the small
$P$ regimes are, in fact, the same one. From a physical point of
view, this provides a physical meaning to the dimensionless parameter
$\gamma$ in the small nonlinearity regime: $\gamma\sim P$. Notice
that, in the general case of a nonlinearity that can be expanded in
power series in $\phi$, the auxiliar nonlinear function $\bar{M}$
can depend itself on $\gamma$. However, for our purposes, it is enough
to consider that $M=O(\gamma)$ ( and thus $\bar{M}=O(1)$) because
we are going to be interested in leading order terms. 

Following the reasoning above, we consider the total field amplitude
as the sum of a large forward amplitude and a small backward amplitude:
$\phi=\phi_{F}+\delta\phi_{B}$. Besides, the small backward amplitude
is a function of the nonlinear parameter $\gamma$ --$\delta\phi_{B}=\delta\phi_{B}(\gamma)$--
and it has to verify that $\delta\phi_{B}(0)=0$, since we are assuming
that no backward radiation is present in the absence of nonlinearities.
The nonlinear function can be then expanded as $M(\phi_{F}+\delta\phi_{B})=M(\phi_{F})+\left(\partial M/\partial\phi|_{\phi_{F}}\right)\delta\phi_{B}+O(\delta\phi_{B})^{2}$.
Substituing the total amplitude $\phi=\phi_{F}+\delta\phi_{B}$ and
the expansion of $M$ in Eq.(\ref{eq:backward_equation_spatial})
and keeping the leading order terms in $\delta\phi_{B}$ only, one
gets\begin{eqnarray*}
\left(-i\frac{\partial}{\partial z}+L_{0}^{1/2}+\frac{1}{2}L_{0}^{-1/2}\left(M(\phi_{F})+\phi_{F}\frac{\partial M}{\partial\phi_{F}}\right)\right)\delta\phi_{B}\\
=-\frac{1}{2}L_{0}^{-1/2}M(\phi_{F})\phi_{F}+O(\delta\phi_{B})^{2}.\end{eqnarray*}
Since $M=O(\gamma)$, we can find the leading order term in $\gamma$
for $\delta\phi_{B}$ from the previous equation:\[
\delta\phi_{B}=-\frac{1}{2}\left(-i\frac{\partial}{\partial z}+L_{0}^{1/2}\right)^{-1}L_{0}^{-1/2}M(\phi_{F})\phi_{F}+O(\gamma)^{2},\]
so that $\delta\phi_{B}$ is also $O(\gamma)$. We can proceed analogously
with the forward FBE (\ref{eq:forward_equation_spatial}) to find\begin{eqnarray*}
\lefteqn{\left(-i\frac{\partial}{\partial z}-L_{0}^{1/2}\right)\phi_{F}=}\\
 &  & \frac{1}{2}L_{0}^{-1/2}M(\phi_{F})\phi_{F}+\frac{1}{2}L_{0}^{-1/2}\left(M(\phi_{F})+\phi_{F}\frac{\partial M}{\partial\phi_{F}}\right)\delta\phi_{B}\\
 &  & +O(\delta\phi_{B})^{2},\end{eqnarray*}
which, after introducing $\gamma$ dependences ($M=O(\gamma)$, $\delta\phi_{B}=O(\gamma)$),
becomes\begin{equation}
\left(-i\frac{\partial}{\partial z}-L_{0}^{1/2}\right)\phi_{F}=\frac{1}{2}L_{0}^{-1/2}M(\phi_{F})\phi_{F}+O(\gamma)^{2}.\label{eq:pure_forward_equation}\end{equation}

Therefore, neglecting backward components is equivalent to consider
FBEs up to $O(\gamma)^{2}$ terms. Or, equivalently, the pure forward
equation (that is, FBEs with $\phi\approx\phi_{F}$ and $\phi_{B}\approx0$)
is just a weak-nonlinearity approximation. More specificaly, it is
the leading-order contribution in the nonlinear parameter $\gamma$
to FBEs. For this reason, it is possible to find equivalent forms
to the forward equation different than (\ref{eq:pure_forward_equation}).
An interesting alternative version of the forward equation is easily
obtained by using the following property\begin{eqnarray*}
\lefteqn{L_{0}^{1/2}\phi_{F}+\frac{1}{2}L_{0}^{-1/2}M(\phi_{F})\phi_{F}+O(\gamma)^{2}=}\\
 &  & L_{0}^{1/2}\left(1+\frac{1}{2}L_{0}^{-1}M(\phi_{F})\right)\phi_{F}+O(\gamma)^{2}=\\
 &  & \left(L_{0}+M(\phi_{F})\right)^{1/2}\phi_{F}+O(\gamma)^{2},\end{eqnarray*}
which allows us to write Eq.(\ref{eq:pure_forward_equation}) as\begin{equation}
-i\frac{\partial}{\partial z}\phi_{F}=\left(L_{0}+M(\phi_{F})\right)^{1/2}\phi_{F}+O(\gamma)^{2}.\label{eq:pure_forward_equation2}\end{equation}

The previous version of the forward equation has been used to simulate
monochromatic nonlinear propagation of spatial structures in photonic
crystal fibers \cite{ferrando-oe11_452} in the non-paraxial regime.
Despite the pure forward equation is first-order in $z$, it has an
intrinsic non-paraxial nature. Unlike in the nonlinear Schr\"odinger
equation, the standard integral $\int\phi_{F}^{*}\phi_{F}$ is not
the conserved quantity. This can be clearly seen if one writes the
evolution operator associated to Eq.(\ref{eq:pure_forward_equation2})
for an infinitesimal axial step $\epsilon$:\[
\phi_{F}(z+\epsilon)=\exp i\epsilon\left(L_{0}+M(\phi_{F}(z))\right)^{1/2}\phi_{F}(z).\]
This evolution operator is not unitary because the operator $L_{0}+M(\phi_{F}(z))$,
despite it is self-adjoint, is not positive-definite, inasmuch as
it can have negative eigenvalues corresponding to evanescent waves
($\beta^{2}<0$). The loss of unitarity is due to this evanescent
modes leading to the non-conservation of the integral $\int\phi_{F}^{*}\phi_{F}$.

\section{Conclusions}

The experimental availability of high nonlinearities is expected to
unveil a number of new effects that will force us to extract information
from Maxwell's equations in a more accurate manner. In this paper
we shed some light into the problem of revealing the close interplay
between backward components and nonlinearities in axially-invariant
systems. With a minimum amount of approximations, we have been able
to find a system of two coupled first-order equations for the forward
and backward spectral components of the electromagnetic field, the
so-called forward-backward equations. The explicit appearence of forward
and backward components as well as of their nonlinear couplings in
these equations is useful to quantify under which conditions nonlinearly-generated
backward components can be relevant in a new scenario of highly-nonlinear
effects. 

From the formal point of view, the FBEs are specially appealing in
the sense that they admit a simple bi-spinor representation that closely
resemble that of a Dirac equation for the positive and negative components
of a fermion wave function in 1+1 dimensions. Similarly to Dirac equation,
the use of the algebraic properties of the spinor FBEs allows us to
obtain the conserved quantities associated to them in an elegant way.
In the same manner, all conserved quantities also admit an interpretation
as conserved charges associated to phase symmetries.

The dimensional reduction is a remarkable issue of the modal approach
followed here. The original 3+1 dimensions (3 spatial, 1 frequency)
of the starting wave-equation for $\mathbf{E}_{\omega}(x,y,z)$ (Eq.(\ref{eq:general_wave_equation}))
are reduced to 1+1 (1 spatial, 1 frequency) in the FBEs. The modal
approach is a way of {}``integrating out'' the transverse spatial
degrees of freedom ($x$ and $y$). Of course, the coupling between
tranverse degrees of freedom in Eq.(\ref{eq:general_wave_equation})
does not dissapear in FBEs. It transforms into the couplings between
the different modal components of the spectral function $c_{n}$ which
are mathematically encoded in the nonlinear matrix function $M_{nn'}$.
For the case in which only a few modal components are relevant, the
process of dimensional reduction provides a dramatical simplification.
Typical envelope equations for propagating pulses are the result of
a similar process in which the propagation of only one linear mode
(usually, the fundamental mode) is assumed. The FBEs, however, provide
a natural way of dealing with a more complex modal structure and,
moreover, they allow one to directly work with the frequency content
of the propagating field; that is, with its spectral components $c_{n}(\omega,z)$.
In this sense, there is no need to resort to the concept of pulse
envelope (unless one wants to make contact to other approaches). For
the same reason, the FBEs are equally suitable to describe pulse propagation
with extremely large bandwiths since no assumption on the form of
$c_{n}(\omega,z)$ is required. Ideally, they could handle any type
of temporal-spectral behavior provided the only assumption needed,
the {}``scalar approximation'' (in its strong or weak form ---see
section \ref{sec:Modal-second-order-equation}---), is reasonably
fulfilled.

Another scenario in which these equations can be useful is that in
which there are relevant spatial-temporal effects. The FBEs inherently
include couplings between frequency and modal indices through the
nonlinear matrix function $M_{nn'}(\omega,\omega';c)$. Since modal
indices correspond to spatial amplitudes of linear modes, the nonlinear
matrix function $M_{nn'}(\omega,\omega';c)$ is constructed out of
these amplitudes as an overlapping integral (section \ref{sec:Modal-second-order-equation}).
Thus, part of the frequency dependence of $M_{nn'}$ is due to the
explicit dependence of linear mode amplitudes on frequency. In the
ultra-wide spectrum regime, this dependence cannot be neglected and,
therefore, the contribution of several spatial modes can produce simultaneous
couplings between modal indices and frequencies which can be naturally
treated in the framework of the FBEs. In terms of the original 3+1
Maxwell's equations, these effects would correspond to spatial-temporal
phenomena, in which the spatial and temporal degrees of freedom of
the electric field $\mathbf{E}(x,y,x,t)$ could not be factorized.

Sumarizing, the FBEs provide a general framework to deal with nonlinearities
in axially-invariant inhomogenous dielectric media, limited only to
a reasonable validity of the {}``scalar approximation'' (in its
strong or weak form). As we have seen in section \ref{sec:Particular-cases},
its generality can be made evident after analizing two apparently
disconnected cases: forward pulse propagation (a purely temporal phenomenon)
and monochromatic non-paraxial evolution of spatial structures (a
purely spatial phenomenon). Both are equally and naturally described
within the FBEs formalism. The fast progress in nonlinear optics experiments
provides effects of increasing complexity both in the spatial and
time domains as well as in the interplay between forward and backward
components. Strong correlations between spatial and time domains and
forward and backward components will play a more and more important
role. In this context, the FBEs can be a suitable and convenient tool
to encompass a variety of different phenomena within a common framework.

\bibliography{FBE}
\end{document}